\documentclass{evn2004}
\usepackage{txfonts}
\usepackage[dvips]{graphicx}
\usepackage[ansinew]{inputenc}
\usepackage[T1]{fontenc}
\usepackage{amssymb}
\begin{document}
   \title{An extremely curved relativistic jet in PKS 2136+141}

   \author{T. Savolainen\inst{1}, K. Wiik\inst{1,2}, E. Valtaoja\inst{1}
          \and
          M. Tornikoski\inst{3}
          }

   \institute{Tuorla Observatory, University of Turku, V\"ais\"al\"antie 20, 
       FI-21500, Piikki\"o, Finland
         \and
	 Institute of Space and Astronautical Science, Japan Aerospace
	 Exploration Agency, 3-1-1 Yoshinodai,
	 Sagamihara, Kanagawa 229-8510, Japan
	 \and
         Mets\"ahovi Radio Observatory, Helsinki University of Technology,
	 Mets\"ahovintie 114, FI-02540, Kylm\"al\"a, Finland
             }

   \abstract{
We report the discovery of an extremely curved jet in the high frequency
peaking GPS quasar PKS 2136+141. Our multi-frequency VLBA images show
a bending jet making a turn-around of 210 degrees on the plane of the sky,
which is, to our knowledge, the largest ever observed change in a structural 
position angle of an extragalactic jet. Images taken at six different
frequencies, from 2 to 43 GHz, beautifully reveal a spiral-like trajectory. We
discuss possibilities to constrain the 3-D geometry of the source and
suggest that it could be used as a testbed for models describing the bending
of the relativistic jets.
   }

   \maketitle

\section{Introduction}

A significant fraction of extragalactic jets show some degree of bending -- 
from slightly curved jets up to a complete turn-around of an almost
180$\degr$. Very large-angle misalignments have been observed in some 
core-dominated radio sources, but, generally, misalignment angles larger than
120$\degr$ are rare (Wilkinson et al. \cite{wil86}, Tingay et
al. \cite{tin98}, Lister et al. \cite{lis01}). Up to today, the largest
observed $\Delta$P.A. is 177$\degr$ in the gamma-ray blazar PKS 1510-089,
which  shows a jet bending almost directly across our line of sight (Homan et
al. \cite{hom02}). 

Since core-dominated radio sources have jets oriented close to our line of 
sight, all intrinsic variations in the jet trajectories are exaggerated in
projection -- often to a large degree. This implies that rather small
intrinsic bends can manifest themselves as large-angle misalignments between
the jet axes observed on parsec and kiloparsec scales, or as high as 90$\degr$
turns in VLBI images.  

Variety of explanations have been suggested for the observed bends including
ballistic motion of plasmoids ejected from a precessing jet nozzle, ram
pressure due to winds in the intracluster medium, the density gradient in the
transition to the intergalactic medium and growing magnetohydrodynamic
instabilities in the jet propagation. Most likely, different mechanisms work
in different sources. It would be valuable to be able to reliably identify the
reason for bending in individual sources, since the observed properties of the
bend -- correctly interpreted -- can constrain several physical parameters of
the jet and the external medium (see e.g. Hardee \cite{har03}). 

In this paper, we present VLBA images showing that PKS 2136+141 (OX 161), a
quasar type GHz-peaked spectrum source at moderately high redshift of
2.427, has a parsec-scale jet, which appears to bend {\it over} 180$\degr$ on
the plane of the sky, possibly making it the most curved astrophysical jet
ever observed.

In VLA images, PKS 2136+141 is compact, showing no extended emission (Murphy
et al. \cite{mur93}), but has a core-jet morphology in VLBI scales. Both 5 GHz
(Fomalont et al. \cite{fom00}) and 15 GHz (Kellermann et al. \cite{kel98})
VLBA observations reveal a core-dominated source with a short, slightly
bending jet. Kellermann et al. (\cite{kel04}) have measured an apparent
superluminal speed of $\beta_{app}=1.8\pm1.4$ for the brightest moving
component.

The radio-mm spectrum of the source peaks at 8-10 GHz in the quiescent
state and strongly inverted spectrum ($\alpha \ge +0.5$) is seen during
outbursts (Tornikoski et al. \cite{tor01}). This puts PKS 2136+141 into the
class of high frequency peakers. The source is variable at radio frequencies
showing a factor of $\sim 3$ variations in cm-wavelength flux curves with
characteristic time scale of $\sim 5$ years (M. Aller, private
communication). Last strong outburst started around 1998 and the cm-flux was
still rising in 2003, indicating that our observations in 2001 caught the
source during the rising phase of a major flare. 

\section{Observations and results}

   \begin{figure*}
     \centering
     \includegraphics[width=1.0\textwidth]{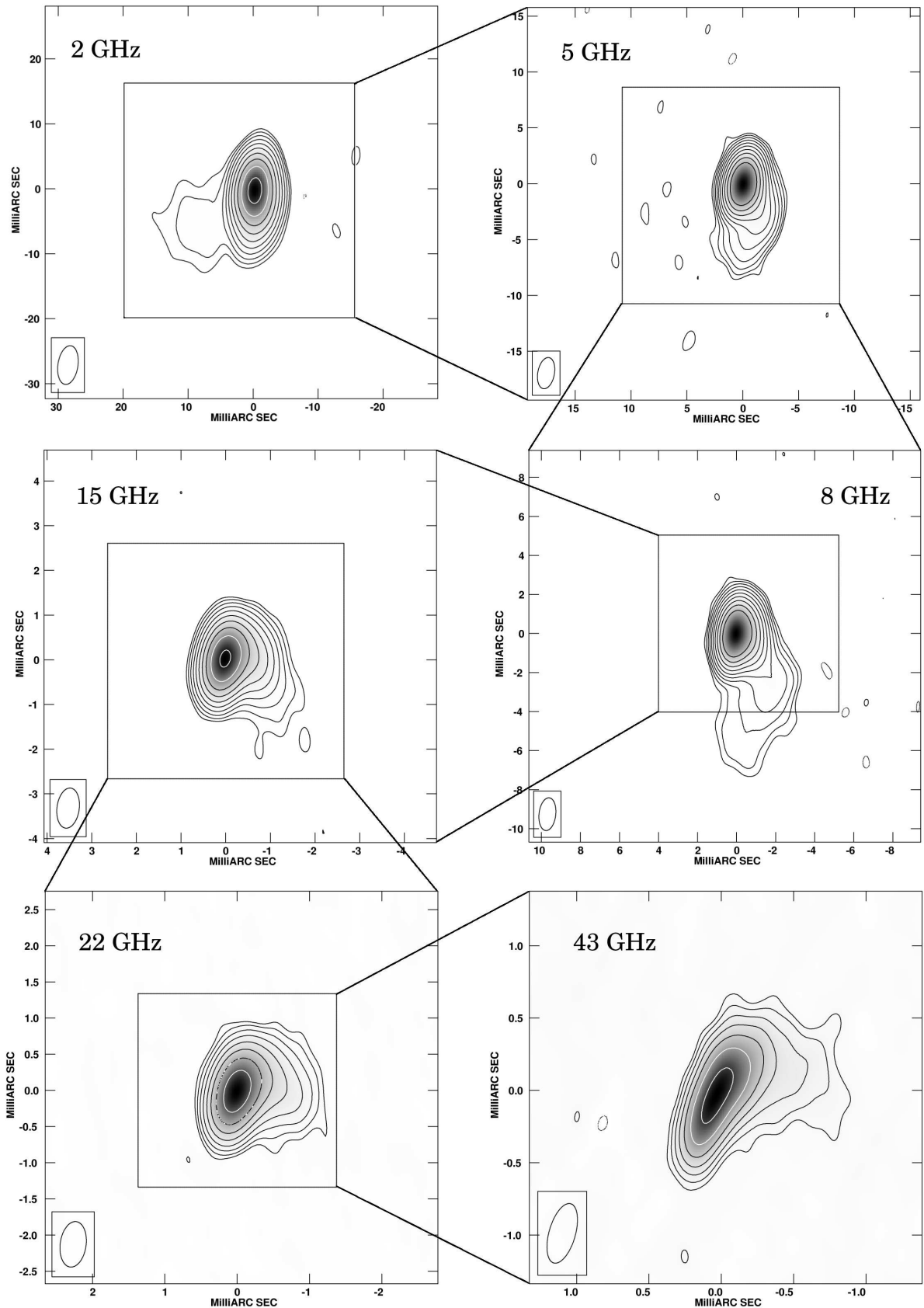}
     \caption{Compiled figure showing the total intensity images at all six 
      frequencies. Both contours and grey scale represent total intensity. Peak
      intensities and contour levels are given in Table~\ref{mapar}.} 
     \label{mutka}
   \end{figure*}

On May 2001 we made multi-frequency polarimetric VLBI observations of four
GPS quasars, including PKS 2136+141, using the National Radio Astronomy 
Observatory's (NRAO) Very Long Baseline Array. Observations were split into 
a high frequency part (15, 22 and 43 GHz), which was observed on May 12th,
and into a low frequency part (2, 5 and 8 GHz) observed on May 14th .
Dual polarization was recorded at all frequencies.

The data were correlated on the VLBA correlator and were postprocessed with
the NRAO's Astronomical Image Processing System, AIPS, (Bridle \& Greisen
\cite{bri94}) and the Caltech DIFMAP package (Shepherd \cite{she97}). Standard
methods for VLBI data reduction and imaging were used. To allow estimation of
the parameters of the emission regions, the self-calibrated visibilities were
model-fitted using the DIFMAP. Circular Gaussian model components were used
and we sought to obtain the best possible fit to the visibilities and to the
closure phases. Several starting points were tried in order to avoid a local
minimum fit. 

Figure~\ref{mutka} displays uniformly weighted CLEAN images of PKS 2136+141 at
all six frequencies, strikingly revealing a jet which gradually bends
$210\degr$ with its structural P.A. turning clockwise from -27$\degr$ at 43
GHz to +123$\degr$ at 2 GHz. The source is rather compact at all frequencies
with maximum jet extent of approximately 15 mas corresponding to $\sim120$ pc
at the source distance\footnote{Throughout the paper we use following values
for cosmological parameters: $H_0$=71km/s/Mpc, $\Omega_M$=0.27 and 
$\Omega_\Lambda $=0.73.}. The core is the brightest component in PKS 2136+141
at all frequencies except 43 GHz, at which the brightness peak is located at
0.28 mas from the core. The brightness profiles along the jet at
frequencies 2--22 GHz are quite smooth without any pronounced knots. At 43 GHz
the core region is either elongated pointing to P.A. of -27$\degr$ or consists
of two distinct components separated by 0.28 mas. 

   \begin{figure*}
     \centering
     \includegraphics[width=1.0\textwidth]{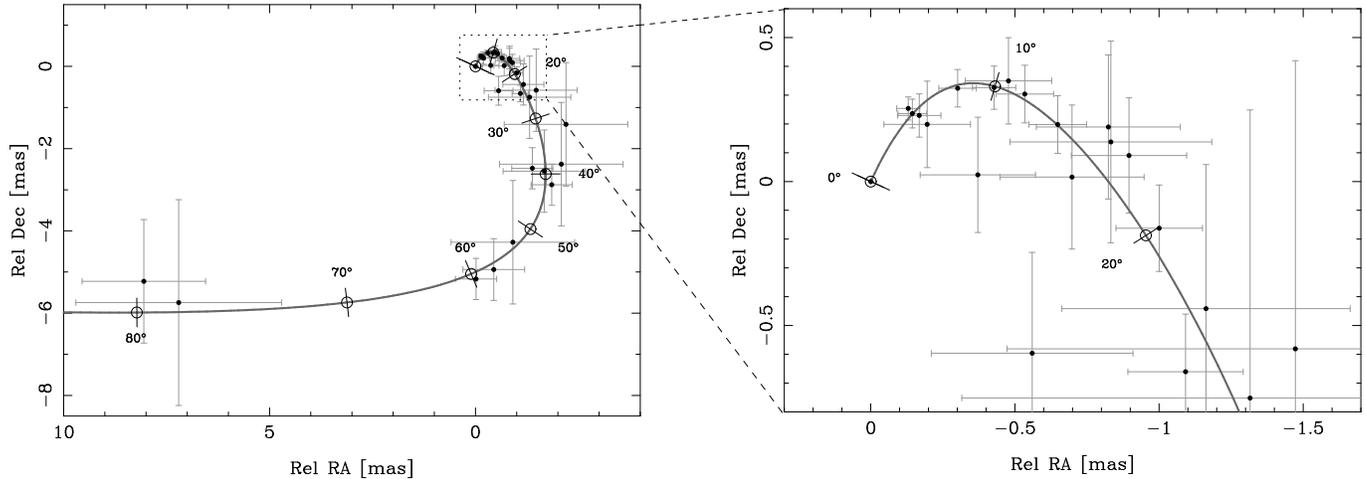}
     \caption{Positions of model-fit components at all six frequencies. At
     each frequency the core is placed to the origin, and hence, no
     frequency shifts of self-absorbed core are taken into account. The
     left image shows the whole source and the right highlights the
     inner 2 mas. The errors in component positions are determined by using
     Difwrap program (Lovell \cite{low00}) and they should be considered as
     very conservative estimates. The solid line represents the best fit of a
     geometrical model describing a  helix on a surface of a cone with
     linearly growing pitch angle.}
     \label{heliksi}
   \end{figure*}

\begin{figure}
     \centering
     \includegraphics[angle=-90,width=0.45\textwidth]{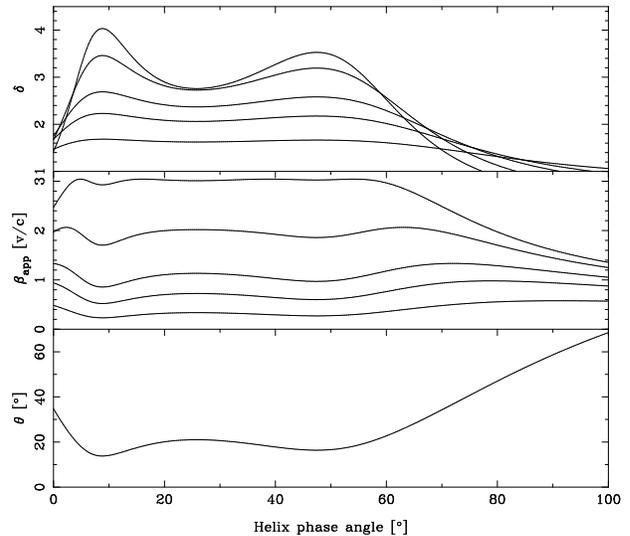}
     \caption{Doppler factor $\delta$, apparent jet speed $\beta_{app}$ and
     angle $\theta$ between the jet and our line of sight as a function of
     phase angle of a helical trajectory depicted in Figure~\ref{heliksi}. The
     assumed jet velocities in two upper panels are $\beta=$ 0.95, 0.9, 0.8,
     0.7 and 0.5 (from top to bottom).} 
     \label{physpar}
   \end{figure}

The positions of model-fitted Gaussian components are represented in
Figure~\ref{heliksi} together with a helical trajectory fitted to the
component positions. In order to see if the spiral-like appearance
of the jet can be described with a helical trajectory, we tried to fit four
simple and purely geometrical models to the data. These included helices on
the surface of the cylinder and on the surface of the cone combined with
either a constant or linearly growing pitch angle. It turned out that only a
helix on the surface of the cone with a linearly increasing pitch angle gives
a satisfactory fit to the data ($\chi^2=1.7$). The parameters describing the
trajectory of the best fit model are: the radius of the cone at the core
position, $R$ = 9.28 mas, the phase angle at the core position,
$\psi$=177.6$\degr$, the half-opening angle of the cone, $\Theta$ =
37.5$\degr$, the growth rate of the pitch angle,
$\mathrm{d}\xi/\mathrm{d}\phi$ = 18.7, and the
angle between the axis of the cone and our line of sight, $\Delta$ =
55.4$\degr$. In Figure~\ref{physpar} we have plotted the Doppler factor,
apparent speed, and the angle between the local direction of the jet and our
line sight as a function of phase angle of our best fitted helical
trajectory. The peak in Doppler factor around phase angle of 10$\degr$ is
almost coincident with apparent brightening of the jet at 0.28 mas from the
core in 43 GHz image. Also, the calculated  $\beta_{app}$ is in agreement with
the value of $1.8\pm1.4$ measured by Kellermann et al. (\cite{kel04}). Thus,
although the model is not based on physical arguments, it fits well to the
observed properties of the source. If the bending is, in fact, due to a
helical trajectory, the helix should be quickly opening and already
asymptotically approaching a straight line in the scale of 15 mas. 
     
\section{Discussion}
\subsection{Reason for bending}
There are several possible mechanisms that could be responsible for bending
of the jet in PKS 2136+141. Projection effects most likely play a role here
and the intrinsic bending angle is much less than the observed one. However,
the mere fact that observed $\Delta$P.A. is larger than 180$\degr$ constrains
the possible mechanisms, since it excludes scenarios where the jet is bent by
a simple change in the density of the external medium, like e.g. deflection by
a massive cloud in the ISM or by a uniform density gradient.  

Homan et al. (\cite{hom02}) explained the large misalignment between the pc
and kpc scale jets of PKS 1510-089 with a scenario where the jet is bent after
it departs the host galaxy, either by the density gradient in the transition
region or by ram pressure due to the winds in the intracluster medium. In PKS
2136+141, the bending takes place well within 15 mas from the core,
corresponding to a linear distance of 120 pc. If we assume a typical scale
size of the elliptical hosts of radio galaxies, $\sim 30$ kpc, the angle
between the jet and our line of sight would have to be considerably less than
$0.2\degr$ if the jet bent after departing the galaxy. This would place our
line of sight inside the jet opening angle, which is not observed. Thus, the
jet in PKS 2136+141 bends before it reaches the outskirts of the host galaxy. 

Another obvious possibility explaining the observed bend is a precessing jet
where the components ejected at different times to different directions move
ballistically, and form an apparently curved locus. Such models have been used
to describe oscillating 'nozzles' observed in some BL Lac sources
(e.g. Stirling et al. \cite{sti03}, Tateyama \& Kingham \cite{tat04}). Jet
precession can be due to the Lense-Thirring effect in case of misalignment
between the angular momenta of accretion disk and a Kerr black hole, or it can
occur in a binary black hole system where secondary black hole tidally induces
precession to the accretion disk. In case of PKS 2136+141, we should be able
to easily test the precessing jet model with future high frequency VLBI
observations. If the jet is ballistic with a precessing inlet, we should
observe the orientation of the most compact part of the jet to change with
time. Also, the components should not follow a curved trajectory, but rather
go along straight lines starting from the core.  

Observed bends in astrophysical jets can also be a manifestation of growing
magnetohydrodynamic instabilities, which can launch helical distortions of the
jet propagating either as a body mode displacing the entire jet or as a
surface mode producing helically twisted patterns on the surface of the jet
(see e.g. Hardee \cite{har87}, \cite{har03}). With current data, it is
difficult to study whether the observed bend in PKS 2136+141 could be modeled
with such instabilities, since no complete 'twist' is observed and thus, the
wavelength, wave speed and growth rate of the disturbance cannot be
unambiguously determined.     

      \begin{table}
	\caption[]{Parameters of Maps}
	\centering
	\label{mapar}  
         \begin{tabular}{cccc}
            \hline
            \noalign{\smallskip}
            Frequency &  rms noise & Peak Intensity & Contour $c_0^{\mathrm{a}}$ \\
	    (GHz) & (mJy beam$^{-1}$) & (mJy beam$^{-1}$) & (mJy) \\
            \noalign{\smallskip}
            \hline
            \noalign{\smallskip}
            2 & 0.6 & 1153 & 1.7 \\
	    5 & 0.4 & 1811 & 1.0 \\
	    8 & 0.3 & 1890 & 0.9 \\
	    15 & 0.7 & 1429 & 2.5 \\
	    22 & 1.9 & 1153 & 5.8 \\
	    43 & 2.2 & 512 & 6.1 \\
            \noalign{\smallskip}
            \hline
         \end{tabular}
\begin{list}{}{}
\item[$^{\mathrm{a}}$] Contour levels are represented by geometric series 
  $c_0(1,...,2^n)$, where $c_0$ is the lowest contour level indicated in 
  the table.
\end{list}
\end{table}

\subsection{3D-trajectory from future observations}
With further VLBI observations, we would be able to measure how the flux
density $S_{obs}$, opening angle $\Psi_{obs}$ and apparent speed $\beta_{app}$
change as a function of both time $t$ and position $z$ along the
jet. $S_{obs}$, $\Psi_{obs}$ and $\beta_{app}$ all depend on the angle
$\theta (z)$ between the local jet direction and our line of sight: 
\begin{eqnarray}
S_{obs}(z,t) & = & [\Gamma(t)(1-\beta(t) \cos{\theta(z)})]^{-(2-\alpha)} \cdot S(z,t) \\
\sin \Psi_{obs}(z) & = & \frac{\tan \Psi(z) \cos \theta(z)}{\sin \theta(z)} \\
\beta_{app}(z,t) & = & \frac{\beta(t) \sin \theta(z)}{1-\beta(t) \cos
  \theta(z)}, 
\end{eqnarray}
where $\beta$ is the flow speed of the jet, $\Gamma = 1/\sqrt{1-\beta^2}$,
$\alpha$ is the measured spectral index ($S\propto \nu ^{\alpha}$), $S(z,t)$
is an intrinsic flux density and $\Psi(z)$ is an intrinsic jet
opening angle. If we take $\beta$ and $\Psi$ to be constants and assume some
simple form for $S(t,z)$, we can try to fit for $\theta(z)$ and thus,
determine the 3-D trajectory of the jet. 

The knowledge about 3-D trajectory would place strong constraints on any
physical model trying to explain the observed curvature in PKS 2136+141,
hopefully making it possible to distinguish between the models. Keeping that
in mind, this source may prove to be a good testbed for different bending
scenarios in future. 

\begin{acknowledgements}
The VLBA is a facility of the National Radio Astronomy Observatory, operated
by Associated Universities, Inc., under cooperative agreement with the U.S. 
National Science Founda\-tion. This research was partly supported by
Finnish Cultural Founda\-tion (T.S.) and the Japan Society for the Promotion of
Science (K.W.). 
\end{acknowledgements}

\end{document}